\documentclass[showpacs,amssymb,aps]{revtex4}
\usepackage{amsmath}
\usepackage{amstext}
\usepackage{amsopn}
\usepackage{amsfonts}
\usepackage{amssymb}
\usepackage{bbm}
\usepackage{accents}
\usepackage{empheq}
\usepackage{graphicx}
\usepackage{epsf}
\usepackage{graphics}
\usepackage[latin1]{inputenc}
\def\sc{\scriptscriptstyle}

\begin{document}

\title{The thermal chiral anomaly in the Schwinger model}

\author{Ashok Das$^{a,b}$ and J. Frenkel$^{c}$\footnote{$\ $ e-mail: das@pas.rochester.edu,  jfrenkel@fma.if.usp.br}}
\affiliation{$^a$ Department of Physics and Astronomy, University of Rochester, Rochester, NY 14627-0171, USA}
\affiliation{$^b$ Saha Institute of Nuclear Physics, 1/AF Bidhannagar, Calcutta 700064, India}
\affiliation{$^{c}$ Instituto de Física, Universidade de São Paulo, 05508-090, São Paulo, SP, BRAZIL}

\begin{abstract}
In the Schwinger model at finite temperature, we derive a closed form result for the chiral anomaly which arises from the long distance behavior of the electric field \cite{frenkel}. We discuss the general properties associated with this thermal anomaly as well as its relation with the ``index" of the Dirac operator. We further show that the thermal anomaly, like the zero temperature anomaly which arises from the ultraviolet behavior of the theory, does not receive any contribution from higher loops. Finally, we determine the complete effective action as well as the anomaly functional on both the thermal branches in the closed time path formalism.
\end{abstract}


\maketitle
\bigskip

In a recent paper \cite{frenkel} we showed that the more pronounced infrared behavior  in a massless gauge theory at finite temperature can lead to a temperature dependent correction to the chiral anomaly for background fields that do not vanish asymptotically. Furthermore, since this (infrared) thermal anomaly does not depend on the ultraviolet behavior of an amplitude, unlike the zero temperature chiral anomaly, it manifests in higher point amplitudes as well. In this paper, we study further the structure of this thermal anomaly and bring out several interesting features associated with it.

Let us recall from \cite{adilson} that in the $1+1$ dimensional Schwinger model \cite{schwinger}, the temperature dependent part of the effective action on the $C_{(+)}$ branch of the thermal contour (in the closed time path formalism \cite{schwinger1,das}) has the form
\begin{align}
& \Gamma_{\rm eff}^{(\beta)} = \sum_{n=1}^{\infty} \Gamma_{2n}^{(\beta)} [\bar{u}\cdot A],\nonumber\\ 
& \Gamma_{2n}^{(\beta)} [\bar{u}\cdot A]  = \frac{1}{(2n)!} \left[\left(\prod_{j=1}^{2n}\int \frac{d^{2}p_{j}}{(2\pi)^{2}}\,(\bar{u}\cdot A)(p_{j})\delta (p_{j,+})\right) \delta^{2}\left(\sum_{i=1}^{2n} p_{i}\right) I_{2n}^{(\beta)} + p_{j,+}\rightarrow p_{j,-}\right].\label{effaction}
\end{align}
Here $\beta =\frac{1}{kT}$ with $T$ denoting the temperature ($k$ Boltzmann constant), $p_{j,\pm}$ correspond to the light-cone components of the momenta and in the rest frame of the heat bath (see ref. \cite{frenkel,adilson,frenkel0} for notations)  
\begin{equation}
\bar{u}^{\mu} (p) = \frac{\epsilon^{\mu\nu}p_{\nu}}{p_{1}}.\label{ubar}
\end{equation}
(Although we should denote the fields on the $C_{(+)}$ branch as $A_{(+)}$, we are ignoring the subscript here for simplicity and will return to this at the end of the paper.) The temperature dependence in the effective action is contained in the factors $I_{2n}^{(\beta)}$ which involve the distribution functions. In the high temperature limit, the leading contribution coming from the factor $I_{2n}^{(\beta)}$ is linear in temperature for all $n$. In this case, the temperature dependent anomaly functional in the Schwinger model can be calculated from \eqref{effaction} and has the simple explicit form on the $C_{(+)}$ thermal branch given by \cite{frenkel} 
\begin{equation}
P_{\mu} J^{\mu (\beta)}_{5} (P) = A^{(\beta)} (P) = \sum_{n=1}^{\infty} A^{(\beta)}_{2n-1} (P),\label{momanomaly}
\end{equation}
where  
\begin{align}
A_{2n-1}^{(\beta)} (P) & = - \frac{(2ie)^{2n} C_{2n}}{2 (2\pi)^{2}}\bigg[\int d^{2}p_{2n-1} \left(\prod_{j=1}^{2(n-1)} \left(d^{2}p_{j} (\bar{u}\cdot A (p_{j})) \delta (p_{j, +})\right)\right)\notag\\
& \quad \times p_{2n-1,+} \delta (p_{2n-1,+}) (\bar{u}\cdot A (p_{2n-1}))\delta^{2} (P-\sum_{i=1}^{2n-1} p_{i}) + p_{j, +}\rightarrow p_{j, -} + {\rm permutations}\bigg]T,\label{companomaly}
\end{align}
with $C_{2n} = \sum\limits_{m=1}^{n} \frac{1}{2m-1}$. 
The temperature dependent anomaly can be written in the coordinate space as$^{\protect{\footnotemark[1]}}$\footnotetext[1]{Equation (19) in ref. \cite{frenkel} has a typo in the sense that $J(x^{\pm}) \leftrightarrow J (x^{\mp})$.}
\begin{equation}
A_{2n-1}^{(\beta)} (x) = - (ie)^{2n} \pi^{2n-3} (2n-1) C_{2n}\left[\left(I_{+} (x)\right)^{2(n-1)} J (x^{-}) + \left(I_{-} (x)\right)^{2(n-1)} J (x^{+})\right] T = A_{2n-1,+}^{(\beta)} (x) + A_{2n-1,-}^{(\beta)} (x),\label{2npoint}
\end{equation}
where (at finite temperature products of singular functions such as the delta function require a regularized definition)
\begin{equation}
I_{\pm} (x) = I (x^{\pm}) =  \int d^{2}y\, {\rm sgn} (x^{\pm} - y^{\pm}) E (y),\label{I}
\end{equation}
with $E (x)$ denoting the electric field and
\begin{equation}
J (x^{+}) = \int dy^{+}\,{\rm sgn} (x^{+}-y^{+}) \left(E(y^{+},\infty) - E (y^{+},-\infty)\right),\ 
J (x^{-})  = \int dy^{-}\,{\rm sgn} (x^{-}-y^{-}) \left(E(\infty,y^{-}) - E(-\infty, y^{-})\right).\label{J}
\end{equation}
In particular, for a background electric field of the form 
\begin{equation}
E (x^{+},x^{-}) = E_{1}\, {\rm sgn} (x^{+}) \delta (x^{-}),\label{model}
\end{equation}
we have shown \cite{frenkel} that 
\begin{equation}
A_{2n-1}^{(\beta)} (x) = - (2E_{1})^{2n-1} (ie)^{2n} \pi^{2n-3} (2n-1) C_{2n}\,|x^{+}|^{2(n-1)}\,{\rm sgn}(x^{-})\, T.\label{example}
\end{equation}

It is clear from the general structure in \eqref{2npoint} as well as the expression \eqref{example} for the particular background that the thermal anomaly exhibits a divergent structure as we go to higher point amplitudes. In fact, \eqref{2npoint} appears to be well behaved only for $|e\pi I (x^{\pm})| < 1$ as was pointed out in \cite{frenkel}. However, as we will show next, the complete anomaly functional 
\begin{equation}
A^{(\beta)} (x) = A_{+}^{(\beta)} (x) + A_{-}^{(\beta)} (x) = \sum_{n=1}^{\infty} A_{2n-1, +}^{(\beta)} (x) + \sum_{n=1}^{\infty} A_{2n-1, -}^{(\beta)} (x),\label{complete}
\end{equation}
is well behaved and this divergent behavior is only a consequence of the perturbative expansion of the thermal anomaly. 

To sum the anomaly functional \eqref{complete} (see also \eqref{2npoint}), let us recall from the standard tables \cite{GR} that
\begin{equation}
C_{2n} = \sum_{m=1}^{n} \frac{1}{2m-1} = \frac{1}{2}\left(C + \ln n\right) + \ln 2 + \frac{B_{2}}{8n^{2}} + \frac{7 B_{4}}{64n^{4}} + \cdots,\  B_{2} = \frac{1}{6}, B_{4} = - \frac{1}{30},\cdots,\label{c2n}
\end{equation}
where $C\simeq 0.577$ denotes the Euler constant. We note that the contributions from terms with the Bernoulli numbers  $B_{2n}$ are suppressed by rapidly increasing constants as well as powers of $n$ in the denominator. In fact, even for $n=1$, it can be checked that the first three terms in \eqref{c2n} give the exact result up to $2\%$ and the accuracy increases rapidly as $n$ increases. Furthermore, we note that the polylogarithm \cite{GR} has a power series expansion for any complex value of $s$ as
\begin{equation}
Li_{s} (z) = \sum_{n=1}^{\infty} \frac{z^{n}}{n^{s}},\label{polylog}
\end{equation}
which holds for $|z|< 1$. However, through analytic continuation, the polylogarithm is a well behaved function for all finite values of $z$ except possibly for a branch point and poles at $ z=1$. It has the integral representation
\begin{equation}
Li_{s+1} (z) = \int\limits_{0}^{z} dt\ \frac{Li_{s} (t)}{t},\quad z \frac{d Li_{s+1} (z)}{dz} = Li_{s} (z),\quad Li_{1} (z) = - \ln (1-z).\label{intrep}
\end{equation}

With this background, let us look at the first term in \eqref{complete} (the sum for the other term will be similar). We note that
\begin{equation}
S_{+} = \sum_{n=1}^{\infty} (2n-1) \left(ie\pi I (x^{+})\right)^{2(n-1)} C_{2n} = \sum_{n=1}^{\infty} (n-\frac{1}{2}) z_{+}^{n-1} \left((C + \ln n) + 2 \ln 2 + \frac{B_{2}}{4n^{2}} + \frac{7B_{4}}{32 n^{4}} + \cdots\right),\label{splus}
\end{equation}
where we have denoted
\begin{equation}
z_{+} = \left(ie\pi I (x^{+})\right)^{2} = -e^{2}\pi^{2} (I (x^{+}))^{2}.\label{z}
\end{equation}
Using \eqref{polylog}, each of the terms in \eqref{splus} can be evaluated to lead to
\begin{equation}
S_{+} = \frac{1}{2}\left[\left(C + 2 \ln 2\right) \frac{1+z_{+}}{(1-z_{+})^{2}} + \frac{1}{z_{+}}\left(\frac{dLi_{s} (z_{+})}{ds}\Big|_{s=0} - 2 \frac{dLi_{s} (z_{+})}{ds}\Big|_{s=-1}\right) + \frac{B_{2}}{4z_{+}}\left(2 Li_{1} (z_{+}) - Li_{2} (z_{+})\right) + \cdots \right].\label{splus1}
\end{equation}
The higher order terms (involving higher Bernoulli numbers) can also be evaluated trivially, but as we have indicated earlier, these lead to negligible contributions to the sum. Each term in \eqref{splus1} (including the first term as well as the neglected higher order terms) can be written in terms of polylogarithms and their first derivative. From \eqref{2npoint} and \eqref{complete} it follows that the complete temperature dependent anomaly at high temperature can be written as 
\begin{equation}
A^{(\beta)} = A_{+}^{(\beta)} + A_{-}^{(\beta)},\quad A_{+}^{(\beta)} = - \frac{(ie)^{2} T}{\pi} S_{+} J (x^{-}),\quad A_{-}^{(\beta)} = - \frac{(ie)^{2} T}{\pi} S_{-} J (x^{+}),\label{aplus}
\end{equation}
where $S_{-}$ can be obtained in a similar manner and has a form similar to \eqref{splus1} with
\begin{equation}
z_{+}\rightarrow z_{-} = - e^{2}\pi^{2} (I (x^{-}))^{2}.\label{zminus}
\end{equation}

We note from \eqref{splus1} and \eqref{aplus}, as well as from the properties of the polylogarithm discussed above, that $A_{\pm}^{(\beta)}$ in the complete anomaly functional are independently well behaved for both large and small values of $z_{\pm}$ except for singularities at $z_{\pm}= 1$. On the other hand, from the definitions in \eqref{z} and \eqref{zminus}, we recognize that the points $z_{\pm}=1$ are outside the physical region and consequently each of $A_{\pm}^{(\beta)}$ in the complete anomaly functional has no singularity at all. In fact, from the definition of the polylogarithm in \eqref{intrep} we note that $\frac{Li_{s} (z)}{z}$ is well behaved even at infinity so that the complete anomaly functional is absolutely well behaved. We point out here that the points $|z_{\pm}| = 1$ can be thought of as the boundary between the weak and the strong coupling. The weak coupling expansion ($|z_{+}| < 1$) of $A_{+}^{(\beta)}$, for example, in \eqref{aplus} would lead to the perturbative calculation of $A_{+}^{(\beta)}$ in \eqref{2npoint}, but it would be incorrect to draw any conclusion about the divergence structure of the anomaly from the perturbative result. Therefore, this provides a simple example of how conclusions about the divergence structure of the perturbation series in a quantum field theory can be erroneous and very different from the actual behavior of the complete quantity. 

Besides the divergence structure of perturbation theory, we can also study from \eqref{aplus} the conditions under which a nontrivial finite thermal anomaly may arise and the consequences following from such an anomaly. Clearly, this depends on the the behavior of the functions $J (x^{\pm})$ defined in \eqref{J}.  As we had already pointed out in \cite{frenkel}, it is clear from the structures in \eqref{J} that the anomaly will be nontrivial if the background electric field satisfies
\begin{equation}
\Delta_{+} E (x^{+}) = E (x^{+}, \infty) - E (x^{+}, -\infty) \neq 0,\quad \text{and/or}\quad \Delta_{-} E (x^{-}) = E(\infty, x^{-}) - E (-\infty, x^{-}) \neq 0,\label{condition1}
\end{equation}
leading respectively to a nontrivial $J (x^{+})$ and/or $J (x^{-})$. In general, we note that, if both $\Delta_{+} E (x^{+})$ and $\Delta_{-} E (x^{-})$ are nontrivial, they must be integrable for the anomaly to be finite. We note that for the special class of electric fields which are separable of the form
\begin{equation}
E (x^{+}, x^{-}) = f (x^{+}) g (x^{-}),\label{specialclass}
\end{equation}
the anomaly takes a simpler form in the sense that if $\Delta_{+} E (x^{+})$ is integrable, it leads to  $\Delta_{-} E (x^{-}) = 0$ and {\em vice versa}. Correspondingly, for this special class of electric fields if $J (x^{+})$ is finite, then $J (x^{-}) = 0$ and {\em vice versa}. The specific model \eqref{model}  (or the other case discussed in \cite{frenkel}), in fact, corresponds to this class of electric fields and the anomaly is finite in this case.

In the presence of an anomaly, the chiral charge is no longer expected to be conserved. In fact, using the two dimensional relation $J^{\mu}_{5} = \epsilon^{\mu\nu} J_{\nu}$, the time evolution of the (thermal) chiral charge is determined to have the form
\begin{equation}
\partial_{t} Q_{5}^{(\beta)} = \int dx\left(A^{(\beta)} (x^{+},x^{-}) + \partial_{x} J_{0}^{(\beta)} (x^{+},x^{-})\right) = \int dx\left(A^{(\beta)} (x^{+},x^{-}) + \frac{1}{2}(\partial_{+} - \partial_{-})  J_{0}^{(\beta)} (x^{+},x^{-})\right),\label{timeevolution}
\end{equation}
which can be integrated to give
\begin{equation}
Q_{5}^{(\beta)} (\infty) - Q_{5}^{(\beta)} (-\infty) = 2\int dx^{+} dx^{-} \left(A^{(\beta)} (x^{+},x^{-}) + \frac{1}{2}(\partial_{+} - \partial_{-})  J_{0}^{(\beta)} (x^{+},x^{-})\right).\label{index}
\end{equation}
We note that the surface term in \eqref{timeevolution} can conventionally be set to zero for asymptotically vanishing fields. However, when fields are nonvanishing asymptotically, as in our case, the surface term is in general nontrivial. Several comments are in order here. The left hand side of \eqref{index} is generally identified with the index of the Dirac operator \cite{atiyah} which is normally identified with the integrated form of the anomaly, but there is an additional contribution in \eqref{index} coming from the long distance behavior of the fields. Furthermore, in the present case, since the fields are nontrivial asymptotically (and the time contour for the thermal field theory lies in a complex plane, see \cite{das}), rotation to Euclidean space as well as defining the theory on a compact manifold is not possible. Correspondingly, it is not clear whether the left hand side of \eqref{index} can still be identified with the index of the Dirac operator. The right hand side is nonzero in general and in fact as we have already argued, the thermal anomaly $A^{(\beta)}$ is linear in $T$ at high temperature and, therefore, defines a continuous function of the temperature. So, even if it is possible to identify the left hand side of \eqref{index} with an index, it will no longer be an integer at finite temperature. This is consistent with the physical reasoning that the left hand side can be written as the difference between the left handed and the right handed fermion numbers and the fermion distributions at finite temperature are continuous functions of the temperature. 

However, there is a special class of background fields for which the right hand side in \eqref{index} vanishes which is worth pointing out.  Let us note from the definition of the axial charge that it is an even function of time for currents that are $CPT$ odd, namely, for
\begin{align}
& (J^{\mu}, J_{5}^{\mu}) (t,x) \xrightarrow{CPT} - (J^{\mu}, J_{5}^{\mu}) (-t,-x) = - (J^{\mu}, J_{5}^{\mu}) (t,x),\nonumber\\
& Q_{5} (-t) = \int dx\, J_{5}^{0} (-t,x) = \int dx\, J_{5}^{0} (-t,-x) = \int dx\, J_{5}^{0} (t,x) = Q_{5} (t).\label{cptinv}
\end{align}
For such currents, it follows that the left hand side of \eqref{index} vanishes. The right hand side can also be seen to vanish from its symmetry properties. The $CPT$ odd currents also imply that the vector potential as well as the electric field are also $CPT$ odd,
\begin{equation}
(A_{\mu}, E) (x^{+},x^{-}) \xrightarrow{CPT} - (A_{\mu}, E) (-x^{+}, -x^{-}) = - (A_{\mu}, E) (x^{+}, x^{-}),
\end{equation}
and we note that the specific model in \eqref{model} has precisely this property.  We have checked explicitly the vanishing of \eqref{index} to the lowest order in this case.

\begin{figure}[ht!]
\includegraphics[scale=.7]{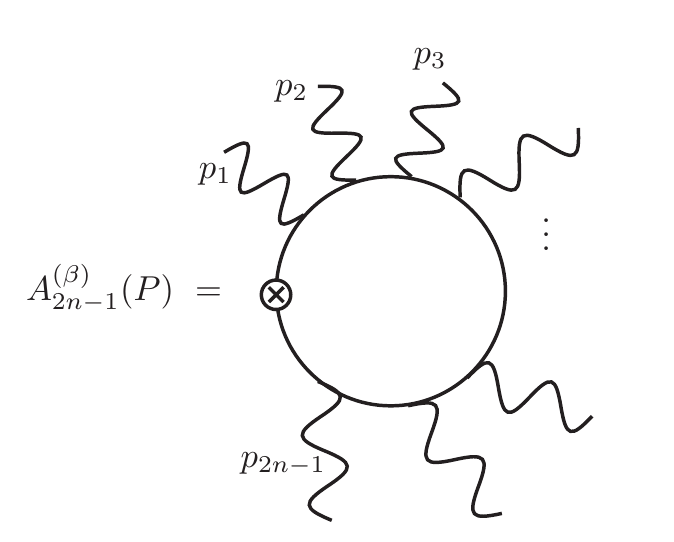}
\caption{The diagram for the $2n$ point anomaly functional $A^{(\beta)}_{2n-1} (P)$ given in \eqref{companomaly}. The circle with a cross denotes the $\gamma_{5}\gamma^{\mu}$ vertex contracted with the momentum $P_{\mu}$.}
\label{1}
\end{figure}

An interesting question that we raised in \cite{frenkel} is whether, like in the case of the conventional anomaly that arises from ultraviolet divergences \cite{adler-bardeen, bardeen,jackiw}, the thermal anomaly also has its contribution only at one loop. In the Schwinger model, this is seen to be true as follows. We note that the standard zero temperature anomaly gives a mass $m^{2} = \frac{e^{2}}{\pi}$ to the photon \cite{schwinger} so that the thermal propagator for the photon (on the $C_{(+)}$ branch in an arbitrary covariant gauge) can be written as
\begin{equation}
D_{\mu\nu} (p) = - \left(\frac{i}{p^{2} - m^{2} + i\epsilon} + 2\pi n_{\sc B} (|p_{0}|) \delta (p^{2}-m^{2})\right) \left(\eta_{\mu\nu} - \frac{\xi p_{\mu}p_{\nu}}{p^{2}}\right),\label{Dmunu}
\end{equation}
where $n_{\sc B} (|p_{0}|)$ denotes the bosonic distribution function and $\xi$ represents the arbitrary gauge fixing parameter. A typical term in the one loop anomaly (in momentum space) given in \eqref{companomaly} can be diagrammatically represented as in Fig. \ref{1}. 
Each external photon line in this diagram comes as a factor $\bar{u}\cdot A (p)$ where the transverse velocity vector $\bar{u}^{\mu} (p)$ satisfies $p\cdot \bar{u} (p) = 0$. Any higher loop correction to the anomaly will arise from one or more pairs of external photon lines in Fig. \ref{1}  being joined to form an internal photon propagator. Every such pairing will lead to a factor of 
\begin{align}
\bar{u}^{\mu} (-p) D_{\mu\nu} (p) \bar{u}^{\nu} (p) \delta (p_{\pm}) & = - \left(\frac{i}{p^{2} - m^{2} + i\epsilon} + 2\pi n_{\sc B} (|p_{0}|) \delta (p^{2}-m^{2})\right)\bar{u}^{\mu} (-p)\bar{u}_{\mu} (p) \delta (p_{\pm})\notag\\
& = - \left(\frac{i}{p^{2} - m^{2} + i\epsilon} + 2\pi n_{\sc B} (|p_{0}|) \delta (p^{2}-m^{2})\right) \left(-\frac{p^{2}}{(p_{1})^{2}}\right) \delta (p_{\pm}) = 0.
\end{align}
Here we have used the transversality of $\bar{u}^{\mu} (p)$ as well as the scalar product following from the  definition in \eqref{ubar}. This shows explicitly that any diagram representing a higher loop correction to the thermal anomaly would vanish so that the one loop anomaly would represent the complete thermal anomaly. There is a rather physical way of understanding this result. Since the photon can be thought of as massive beyond one loop, the photon field (electric field) must necessarily fall off asymptotically. On the other hand, the  (infrared) thermal anomaly arises only if the electric field is nonvanishing asymptotically and therefore, there cannot be any further correction to the thermal anomaly beyond one loop. 

To conclude our discussions on the thermal anomaly, we note that our discussions so far have  been within the context of time ordered Feynman amplitudes restricted to the $C_{(+)}$ thermal branch where the effective action has the explicit form $\Gamma_{\rm eff}^{(\beta)} [\bar{u}\cdot A_{(+)}]$ given in \eqref{effaction}. (Here we have restored the ``$(+)$" thermal subscript which we had ignored throughout the earlier discussions.) It can be checked that the retarded thermal amplitudes, including the retarded anomaly amplitudes, vanish in this theory which can be seen using the forward scattering amplitudes \cite{brandt,brandt1}. Physically, this is a consequence of charge conjugation invariance together with helicity conservation in amplitudes for massless fermions in $1+1$ dimensions. As we have discussed earlier \cite{frenkel1,frenkel2}, vanishing of retarded amplitudes is sufficient to determine the complete effective action (as well as the anomaly functional) on both the thermal branches $C_{(+)}$ and $C_{(-)}$ by generalizing
\begin{equation}
\Gamma_{\rm eff}^{(\beta)} [\bar{u}\cdot A_{(+)}] \rightarrow \Gamma_{\rm eff}^{(\beta)} [\bar{u}\cdot (A_{(+)}-A_{(-)})],
\end{equation}
and similarly for the thermal anomaly. We had already shown such a dependence of the effective action for background fields which are asymptotically vanishing. However, it continues to hold even when the fields may not vanish asymptotically.
 \bigskip

\noindent{\bf Acknowledgments}
\medskip

This work was supported in part  by US DOE Grant number DE-FG 02-91ER40685,  by CNPq and FAPESP (Brazil).

\end{document}